\documentstyle[aps,12pt,preprint,epsfig]{revtex}
\def \beq {\begin{equation}}
\def \eeq {\end{equation}}
\def \ba {\begin{eqnarray}}
\def \ea {\end{eqnarray}}

\def \< {\langle}
\def \> {\rangle}
\draft
\begin{document}

\title{Photoproduction of $\eta _c$ in NRQCD}
\author{Li-Kun Hao and Feng Yuan}
\address{\small {\it Department of Physics, Peking University, Beijing
100871, People's Republic of China}}
\author{Kuang-Ta Chao}
\address{\small {\it China Center of Advanced Science and Technology
(World Laboratory), Beijing 100080, People's Republic of China\\
and Department of Physics, Peking University, Beijing 100871,
People's Republic of China}}
\maketitle

\begin{abstract}
We present a calculation for the photoproduction of 
$\eta_c$ under the framework of NRQCD factorization formalism.
We find a quite unique feature that the color-singlet contribution to
this process vanishes at not only the leading order but also
the next to leading order perturbative QCD calculations
and that the dominant contribution comes from the
color-octet ${}^1S_0^{(8)}$ subprocess.
The nonperturbative color-octet matrix element
of ${}^1S_0^{(8)}$ is related to
that of ${}^3S_1^{(8)}$ of $J/\psi$ by the heavy quark
spin symmetry, and the latter can be from the direct production
of $J/\psi$ at large transverse momentum at the Fermilib Tevatron.
We then conclude that the measurement
of this process may clarify the existing conflict between the color-octet
prediction and the experimental result on the $J/\psi$ photoprodution.
\end{abstract}

Studies of heavy quarkonium production in high energy collisions may provide
important information on both perturbative and nonperturbative QCD.
It is well known from compelling theoretical \cite{GTB,GTB95} and phenomenological 
\cite{FA} reasons that the Color Singlet model (CSM) used to describe the
production and decay of quarkonium is not complete. In the CSM, it is assumed
that the $c\overline{c}$ must be produced in a color-singlet state with the same
angular-momentum quantum number as the charmonium meson which is eventually
observed.
However, it is well know that perturbative QCD calculations of production
and decay of P-wave quarkonia within the CSM are plagued by infrared
divergences \cite{RB}. More recently,
measurements made at the Fermilab Tevatron show that
the CSM also fails to predict the production cross sections of
S-wave quarkonia (e.g. $J/\psi $ and $\psi ^{\prime }$)\cite{FA,AS}. The naive CSM
has been supplanted by the NRQCD factorization formalism \cite{GTB95},
which allows the infrared safe calculation
of inclusive charmonium production and decay rates.
In this approach, the production process is
factorized into short and long distance parts, while the latter is
associated with the nonperturbative matrix elements of four-fermion
operators. 
It also predicts a new
mechanism called the color-octet mechanism in which a $c\overline{c}$ pair is produced
at short distances in a color-octet state, and hadronizes into a final state charmonium
nonperturbatively. This color-octet mechanism might provide an explanation for the
Tevatron data on the surplus of $J/\psi $ and $\psi ^{\prime }$ production\cite{surplus,s1}.

Even though the color-octet mechanism has achieved some successes in describing
the production and decay of heavy quarkonia, more tests of this mechanism are still needed.
Recently, the photoproduction data from HERA \cite{photo} puts
a question about the universality of the color-octet matrix elements
\cite{photo2,MK}, in which the fitted values of the matrix elements 
$\langle {\cal O}^{J/\psi}_8 ({}^1S_0)\rangle$ and $\langle {\cal
O}^{J/\psi}_8({}^3P_J)\rangle$ are one order of magnitude smaller than
those determined from the Tevatron data \cite{s1}.
(More recently, possible
solutions for this problem have been suggested in\cite{explain,explain2}.
In this situation it is certainly helpful to find
other processes to test the color-octet mechanism in the heavy quarkonium
production. 

In this paper, we will calculate photoproduction of $\eta _c$,
$\gamma p\rightarrow \eta_c+X$, under the framework of NRQCD. 
The photoproduction processes proceed predominantly through
photon-gluon fusion at high energies.
In NRQCD the Fock state expansion for $\eta_c$ is
\begin{eqnarray}
\label{zankai}
\nonumber
|\eta_c \rangle
&=&O(1)|c\bar{c}({}^1S_{0},\b{1}) \rangle +
O(v)|c\bar{c}({}^1P_{1},\b{8})g\rangle +
    O(v^2)|c\bar{c}({}^3S_1,\b{8}~ or~ \b{1})gg\rangle\\
   && +O(v^2)|c\bar{c}({}^1S_0,\b{8}~ or~ \b{1})gg\rangle+\cdots.
\end{eqnarray}
This expansion has also been given in Ref.\cite{hadro}. But, they missed
the last term in their expansion.
This term is crucially important for $\eta_c$ production in $\gamma p$
process which will be shown in the following.
For the production of $\eta_c$, the 
contributions to the NRQCD matrix elements 
from the last three terms of the above expansion are the same
order of $v^2$ according to the NRQCD velocity scaling rules. They are all suppressed
by $v^4$ compared to the contribution from the first term (the color-singlet contribution).
However, the color-singlet contributions to $\eta_c$ photoproduction
vanish in the leading order and the
next to leading order $\gamma g$ fusion processes (see the following).
So, the photoproduction of $\eta_c$ is purely a color-octet process,
even to the next to leading order of QCD calculations.
Therefore, the $\eta_c$ photoproduction processes such
as at the HERA, will provide an important test for the color-octet production
mechanism.
Furthermore, we will show by the following calculations, the dominant contribution
to $\eta_c$ photoproduction comes from
the last term of the Fock state expansion in equation
(\ref{zankai}). For this term, the associated color-octet production matrix
element $\langle {\cal O}_8^{\eta_c}({}^1S_0)\rangle $ can be related to the matrix element
$\langle {\cal O}_8^{\psi}({}^3S_1)\rangle $ by the heavy quark symmetry. And the later
color-octet matrix element is important for the color-octet explanation of
the prompt $J/\psi$ surplus production at the Tevatron\cite{surplus,s1}.
So, the measurement of $\eta_c$ photoproduction processes
is closely associated with the 
test of the color-octet gluon fragmentation mechanism proposed in \cite{surplus}.

According to the NRQCD  factorization formalism\cite{GTB95}, the production process
$\gamma+ g\rightarrow \eta_c+X$ can be expressed as the following form,
\begin{equation} \label{xs}
\sigma(\gamma+ g\rightarrow \eta_c+X)=\sum\limits_n F(\gamma+g\rightarrow n+X)\langle {\cal
O}_n^{\eta_c}\rangle .
\end{equation}
Here, $n$ denotes the $c\bar c$ pair configuration in the expansion terms of 
Eq. (\ref{zankai}) (including angular momentum $^{2S+1}L_J$ and color index
1 or 8). $F(\gamma+g\rightarrow n+X)$ is the short distance coefficient for the subprocess
$\gamma +g\rightarrow n+X$. $\langle {\cal O}_n^{\eta_c}\rangle $ is the long distance
non-perturbative matrix element which represents the probability of the
$c \bar c$ pair in $n$ configuration evolving into the physical state
$\eta_c$. The short distance coefficient $F$ can be calculated by using
perturbative QCD in expansion of powers of $\alpha_s$. The long distance
matrix elements are still not available from the first principles at
present. However, the relative importance of the contributions from
different terms in (\ref{xs}) can be estimated by using the NRQCD velocity
scaling rules.

From Eq.(\ref{zankai}), we can see that
the color-singlet contribution to the production of $\eta_c$
is at the leading order in $v^2$.
The associated short distance coefficient is given by the subprocess
\begin{equation}
\gamma g\rightarrow c\bar{c}({}^1S_{0},\b{1}) +g.
\end{equation}
This process occurs at the next to leading order in $\alpha_s$ for the $\gamma g$
fusion processes.
However, there is no contribution from this process,
because it violates $C$(charge) parity conservation.
(The $C$ parities
of the two gluon system (constrained in color-singlet) and the $c\bar c$ pair in
$({}^1S_{0},\b{1})$ state are both $+1$, while the $C$ parity of the photon
is $-1$).

The color-octet contributions to the $\eta_c$ photoproduction come from
the leading order and the next to leading order $\gamma g$ fusion processes
as shown in Fig.1.
At the leading order in $\alpha_s$, the subprocess is $2\rightarrow 1$ (Fig.1(a)),
\begin{equation}
\gamma g\rightarrow c\bar{c}({}^1S_0,\b{8}).
\end{equation}
For this process, we readily have\cite{photo2}
\begin{equation}
\sigma(\gamma+g\rightarrow c\bar c({}^1S_0,\b{8})\rightarrow\eta_c)=\frac{\pi^3e_c^2\alpha\alpha_s}
        {m_c^3}\delta(\hat{s}-4m_c^2)\langle {\cal O}_8^{\eta_c}({}^1S_0)\rangle ,
\end{equation}
where $\hat{s}$ is the invariant mass squared of the partonic process. $m_c$ is
the charm quark mass, and we have approximated the charmonium bound state
mass of $\eta_c$ by $2m_c$.

At the next to leading order in $\alpha_s$, the subprocesses are $2\rightarrow 2$ (Fig.1(b)
and Fig.1(c)). These processes contain the contributions from
\begin{eqnarray}
\gamma +g&\rightarrow& c\bar{c}({}^1S_0,\b{8})+g,\\
\gamma +g&\rightarrow& c\bar{c}({}^3S_1,\b{8})+g,\\
\gamma +g&\rightarrow& c\bar{c}({}^1P_1,\b{8})+g.
\end{eqnarray}
The last two subprocesses only have contributions from diagrams shown in
Fig.1(b). The diagrams shown in Fig.1(c) do not contribute to $({}^3S_1,\b{8})$
and $({}^1P_1,\b{8})$ subprocesses because of $C$ parity
conservation, and they only contribute to the
$({}^1S_0,\b{8})$ production subprocess.
The cross sections for these $2\rightarrow 2$ subprocesses can be expressed
as the following form,
\begin{equation}
\frac{d\sigma}{d\hat{t}}(\gamma+g\rightarrow \eta_c +X)=\frac{1}{16\pi\hat{s}^2}F({}^{2S+1}L_J^{(8)})
\langle {\cal O}_8^{\eta_c}({}^{2S+1}L_J)\rangle ,
\end{equation}
where $\hat{t}=(z-1)\hat{s}$, and $z$ is defined as $z=p\cdot k_{\eta_c}/
p\cdot k_\gamma$ with $p$, $k_{\eta_c}$, $k_{\gamma}$ being the momenta of the proton,
the outgoing $\eta_c$ and the incident photon respectively.
For the short distance coefficients $F$ of the subprocesses
${}^3S_1^{(8)}$ and ${}^1S_0^{(8)}$, we readily have\cite{photo2,photo3}
\begin{eqnarray}
F({}^3S_1^{(8)})&=&
\frac{80(4\pi)^3(2m_c)^2e_c^2\alpha\alpha_s^2}{3( \hat{s}+\hat{t}
) ^2( \hat{s}+\hat{u}) ^2( \hat{t}+
\hat{u}) ^2}(\hat{s}^2\hat{t}^2 + \hat{s}^2
\hat{t}\hat{u} + \hat{s}^2\hat{u}^2 + \hat{s}
\hat{t}^2\hat{u} + \hat{s}\hat{t}\hat{u}^2 + 
\hat{t}^2\hat{u}^2),\\
\nonumber
F({}^1S_0^{(8)})&=&
\frac{6(4\pi)^3\hat{s}\hat{u}\alpha\alpha_s^2}{m_c\hat{t}( 
\hat{s}+\hat{t}) ^2( \hat{s}+\hat{u})
^2( \hat{t}+\hat{u}) ^2}(4\hat{s}^4 + 8\hat{s}
^3\hat{t} + 8\hat{s}^3\hat{u} + 13\hat{s}^2\hat{t}^2
+ 26\hat{s}^2\hat{t}\hat{u} + 13\hat{s}^2\hat{u}^2 \\
&&+
10\hat{s}\hat{t}^3 + 28\hat{s}\hat{t}^2\hat{u} + 26
\hat{s}\hat{t}\hat{u}^2 + 8\hat{s}\hat{u}^3 + 5
\hat{t}^4 + 10\hat{t}^3\hat{u} + 13\hat{t}^2\hat{u}
^2  + 8\hat{t}\hat{u}^3 + 4\hat{u}^4).
\end{eqnarray}
Here, $\hat{s}$, $\hat{t}$ and $\hat{u}$ satisfy the relation
$\hat{s}+\hat{t}+\hat{u}=4m_c^2$.
For the ${}^1P_1^{(8)}$ subprocess, we calculate the short distance
coefficient and get
\begin{eqnarray}
\nonumber
F({}^1P_1^{(8)})&=&
\frac{320(4\pi)^3e_c^2\alpha\alpha_s^2}{3( \hat{s}+\hat{t})
^3( \hat{s}+\hat{u}) ^3( \hat{t}+\hat{u}
) ^3}(\hat{s}^6\hat{t} + \hat{s}^6\hat{u} + 4
\hat{s}^5\hat{t}^2 + 4\hat{s}^5\hat{t}\hat{u} + 4
\hat{s}^5\hat{u}^2 + 7\hat{s}^4\hat{t}^3\\
\nonumber
&& + 11\hat{s}
^4\hat{t}^2\hat{u} + 11\hat{s}^4\hat{t}\hat{u}^2 + 7
\hat{s}^4\hat{u}^3 + 7\hat{s}^3\hat{t}^4 + 14\hat{s}
^3\hat{t}^3\hat{u} + 16\hat{s}^3\hat{t}^2\hat{u}^2 +
14\hat{s}^3\hat{t}\hat{u}^3 \\
\nonumber
&&+ 7\hat{s}^3\hat{u}^4 + 4
\hat{s}^2\hat{t}^5 +11\hat{s}^2\hat{t}^4\hat{u} + 16
\hat{s}^2\hat{t}^3\hat{u}^2 + 16\hat{s}^2\hat{t}^2
\hat{u}^3 + 11\hat{s}^2\hat{t}\hat{u}^4 + 4\hat{s}^2
\hat{u}^5\\
\nonumber
&& + \hat{s}\hat{t}^6 + 4\hat{s}\hat{t}^5
\hat{u} + 11\hat{s}\hat{t}^4\hat{u}^2 + 14\hat{s}
\hat{t}^3\hat{u}^3 + 11\hat{s}\hat{t}^2\hat{u}^4 + 4
\hat{s}\hat{t}\hat{u}^5 + \hat{s}\hat{u}^6 + 
\hat{t}^6\hat{u} \\
&&+ 4\hat{t}^5\hat{u}^2 + 7\hat{t}^4
\hat{u}^3 + 7\hat{t}^3\hat{u}^4 + 4\hat{t}^2\hat{u}
^5  + \hat{t}\hat{u}^6).
\end{eqnarray}

Have provided all the cross section formulas for the subprocesses,
we may estimate the $\eta_c$ production rate in the
photoproduction processes. Because there is no color-singlet contribution
in our calculation, all of the following numerical results are for the
color-octet production.
The results depend on the numerical
values of $\alpha _s$, $m_c$, and the factorization scale $Q$. We use $
\alpha _s( m_c^2) =0.3,~ m_c=1.48GeV$ and $Q^2=( 2m_c)^2 $.
For the parton distribution function of the proton, we use the GRV LO
parametrization\cite{grv}.

In Fig.2, we show the $\eta_c$ photoproduction rate
via the $2\rightarrow 1$ subprocess, in which the intermediate color-octet state is
${}^1S_0^{(8)}$. Following the heavy quark spin symmetry, we estimate the associated
color-octet matrix element $\langle {\cal O}_8^{\eta_c}({}^1S_0)\rangle $ to be
\begin{equation}
\langle {\cal O}_8^{\eta_c}({}^1S_0)\rangle \approx \langle {\cal O}_8^{\psi}({}^3S_1)\rangle =1.06\times 10^{-2} GeV^3.
\end{equation}
The value of the color-octet matrix element $\langle {\cal O}_8^{J/\psi
}({}^3S_1)\rangle$ follows the fitted value in\cite{benek} by comparing the
theoretical prediction of direct $J/\psi$ production to the experimental
data at the Tevatron. The $2\rightarrow 1$ process contributes the $\eta_c$ photoproduction
in the forward direction ($z\sim 1$ and $p_T^2\approx 0$).
In Ref.\cite{photo2}, the authors calculated the $J/\psi$ production
in the forward direction and found some conflicts between the NRQCD prediction
and the experimental measurements.
So, the numerical result shown in Fig.2 may be also questioned by the same
problem.
However, we note that in the case of $J/\psi$ production, the conflicts with
experiment focus on the values of the color-octet matrix elements
$\langle {\cal O}_8^{\psi}({}^1S_0)\rangle $ and $\langle {\cal O}_8^{\psi}({}^3P_0)\rangle $.
which are determined by the lower $p_T$ distribution data of direct
$J/\psi$ production at the Tevatron\cite{s1}.
The lower $p_T$ $J/\psi$ production may be strongly affected
by the initiate and final states gluons radiation, which makes the determination
of the matrix elements
$\langle {\cal O}_8^{\psi}({}^1S_0)\rangle $ and $\langle {\cal O}_8^{\psi}({}^3P_0)\rangle $
from the Tevatron data very poor\cite{explain}.
However, we see that the $\eta_c$ photoproduction in the
forward direction of Fig.2 depends on the value of the color-octet matrix element
$\langle {\cal O}_8^{\eta_c}({}^1S_0)\rangle $, which is related to the color-octet matrix
element $\langle {\cal O}_8^{\psi}({}^3S_1)\rangle $ by the heavy quark symmetry.
And the later color-octet matrix element is determined by the large $p_T$
$J/\psi$ production data at the Tevatron, which can be viewed as a more reliable
estimate because in the large $p_T$ region gluon fragmentation is dominant.

In Fig.3-4, we show the photoproduction cross section of $\eta_c$ via the $2\rightarrow 2$
subprocesses at the HERA, to which there are three color-octet subprocesses
contributions (${}^1S_0^{(8)}$, ${}^3S_1^{(8)}$ and ${}^1P_1^{(8)}$ subprocesses).
In Fig.3, we plot the differential cross section $d\sigma/dz$
distribution for $\eta_c$ production. In Fig.4, we plot $d\sigma/dp_T^2$ distribution.
In these two figures, there are contributions from ${}^3S_1^{(8)}$ and
${}^1P_1^{(8)}$, for which we estimate the associated color-octet matrix elements
by using the naive NRQCD velocity scaling rules,
\begin{eqnarray}
\langle {\cal O}_8^{\eta_c}({}^3S_1)\rangle &\approx &\langle {\cal O}_8^{\eta_c}({}^1S_0)\rangle ,\\
\langle {\cal O}_8^{\eta_c}({}^1P_1)\rangle &\approx &\langle {\cal O}_8^{\eta_c}({}^1S_0)\rangle .
\end{eqnarray}
In these two figures, we impose a cut on $z$ and $P_T^2$:
$0.2<z<0.8$ for Fig.4 and $P_T^2>1GeV^2$ for Fig.3.
From these two figures, we can see that the contribution from the ${}^1S_0^{(8)}$
subprocess dominates over the contributions from the other two processes.
Furthermore, the ${}^1S_0^{(8)}$ contribution to the $z$ distribution
of the differential cross section (Fig.3) rises rapidly as $z$ increases.
We note that for the $J/\psi$ photoproduction process, the $z$ distribution
of the cross section from the color-octet contribution has also this property\cite{photo2}.
However, this is not consistent with the experimental results\cite{photo}.
This conflict between the color-octet model prediction and the experiment
invoke a lot of interesting for the further investigations on the
$J/\psi$ photoproduction\cite{explain,explain2}.
In \cite{explain}, the authors noted that the initiate and final state gluon
radiation may strongly affect the determinations of the color-octet
matrix elements $\langle {\cal O}_8^{\psi}({}^1S_0)\rangle $ and $\langle {\cal O}_8^{\psi}({}^3P_J)\rangle $
from the direct $J/\psi$ production data.
This is because the determinations of these matrix elements are sensitive to the
data of $J/\psi$ production in the lower $p_T$ region, in which the gluon radiation
effects are important.
These matrix elements are also the dominant contributions to the $J/\psi$
photoproduction in the high $z$ region.
So, by involving these effects, they lowered down the determined values of these
color-octet matrix elements, and found they would lead to a consistent
description of the $z$ distribution for the $J/\psi$ photoproduction.
However, these arguments do not affect the $\eta_c$ photoproduction
of our results in Fig.3. This is because the dominant contribution in Fig.3
comes from the color-octet matrix element $\langle {\cal O}_8^{\eta_c}({}^1S_0)\rangle $,
which can be safely related to the matrix element $\langle {\cal O}_8^{\psi}({}^3S_1)\rangle $
by the heavy quark symmetry, and the value of the matrix element
$\langle {\cal O}_8^{\psi}({}^3S_1)\rangle $ we used is from the fitting to the $J/\psi$
production data in the large $p_T$\cite{benek} region which is weakly affected by the
initiate and final states gluon radiation\cite{explain} andtherefore can be viewed as
a reliable estimate.
Another explanation conflict problem emphasizes that the intrinsic $k_T$ may be
important for the $J/\psi$ photoproduction\cite{explain2}. If this is true,
our predictions of Fig.3 and Fig.4 for the $\eta_c$ photoproduction will also
be affected by the intrinsic $k_T$ effects.
Nevertheless, the measurement of the $\eta_c$ photoproduction can provide a consistent
test of these explanations of the $J/\psi$ photoproduction conflicts\cite{explain,explain2}.

For the experimental observation of $\eta_c$, one may choose the following
decay modes, such as $\eta_c\rightarrow \phi\phi$,
$\eta_c\rightarrow \gamma\gamma$ and $\eta_c\rightarrow \rho\rho$,
which either have substantial branching ratios or may be easy to detect.
From Figs.2-4, we can see that the production rate of $\eta_c$ in the photoproduction
processes is comparable with that of $J/\psi$, and we hope that
the experimental measurement of $\eta_c$ photoproduction will soon appeare.

In conclusion, we calculated the $\eta_c$ photoproduction under the NRQCD
factorization formalism. We find that the color-singlet contribution to
this process vanishes at the leading order and the next to leading order
QCD calculations. Therefore, this process is a pure color-octet process, and the measurement
of this process can provide useful information to clarify the existing conflict between the color-octet
prediction and the experimental results on the $J/\psi$ photoprodution.

{\it Acknowledgments}: This work was supported in part by the National Natural Science Foundation
of China, the State Education Commission of China, and the State
Commission of Science and Technology of China.

\newpage
\vskip 10mm
\centerline{\bf \large Figure Captions}
\vskip 1cm
\noindent
FIG. 1. The generic Feynman diagrams for the photoproduction of $\eta_c$
at the leading order $(a)$ and the next to leading order $(b) and (c)$
$\gamma g$ fusion processes. All of these diagrams are for the
color-octet processes.

FIG. 2. The cross section of $\gamma +g\rightarrow\eta_c+X$ in the forward direction
($z\sim 1,~p_T^2\approx 0$) as a function of the photon-proton c.m. energy
$E_{c.m.}=\sqrt{s_{\gamma p}}$. 

\noindent
FIG. 3. The differential cross sections $d\sigma /dz$ for the process $\gamma
+p\rightarrow \eta _c+X$ at the HERA as a function of $z\equiv E_{\eta
_c}/E_\gamma $, where $\sqrt{s_{\gamma p}}=100GeV$.The carves denote contributions from
the color-octet $^3S_1$, $^1P_1$, and $^3S_1$ respectively.

\noindent
FIG. 4. The differential cross sections $d\sigma /dP_T^2$ for $\gamma
+p\rightarrow \eta _c+X$ at the HERA as a function of $p_T^2$,
where $\sqrt{s_{\gamma p}}=100GeV$.The carves denote contributions from
the color-octet $^3S_1$, $^1P_1$, and $^3S_1$ respectively.

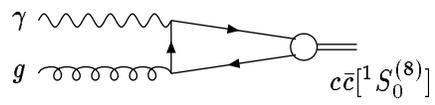
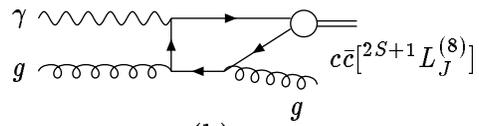
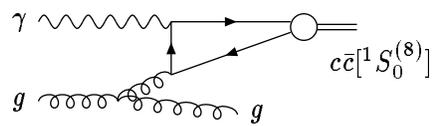
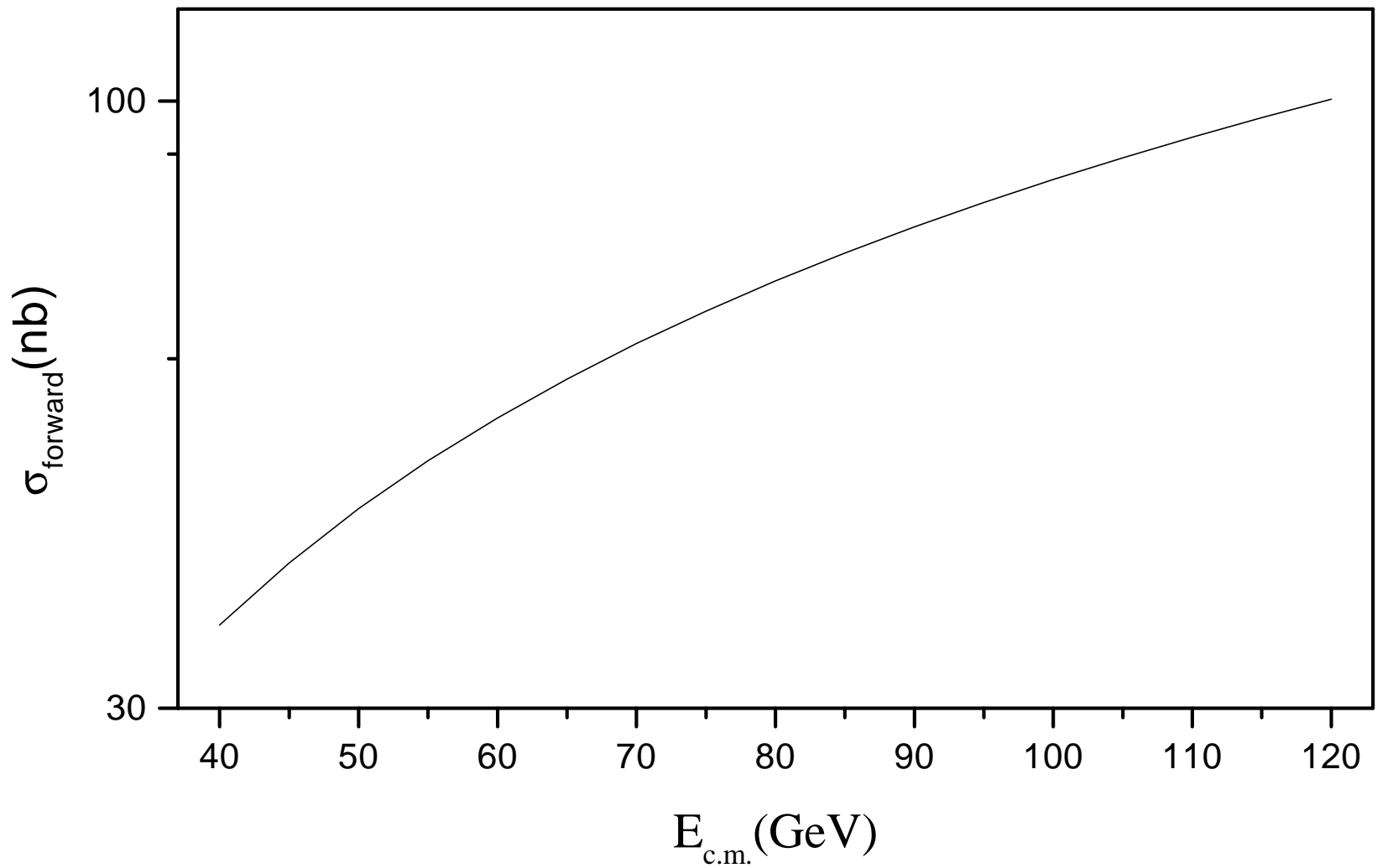
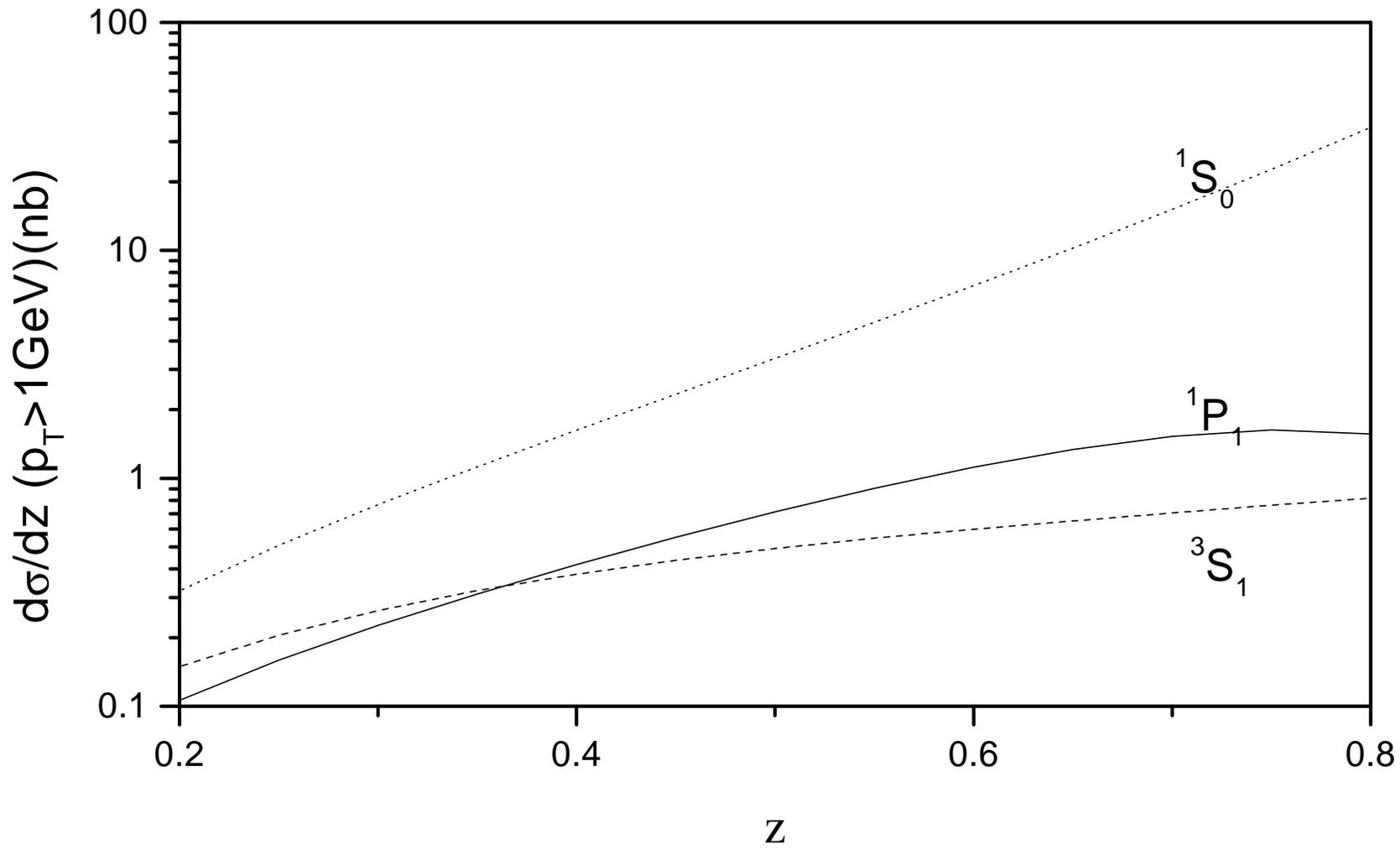
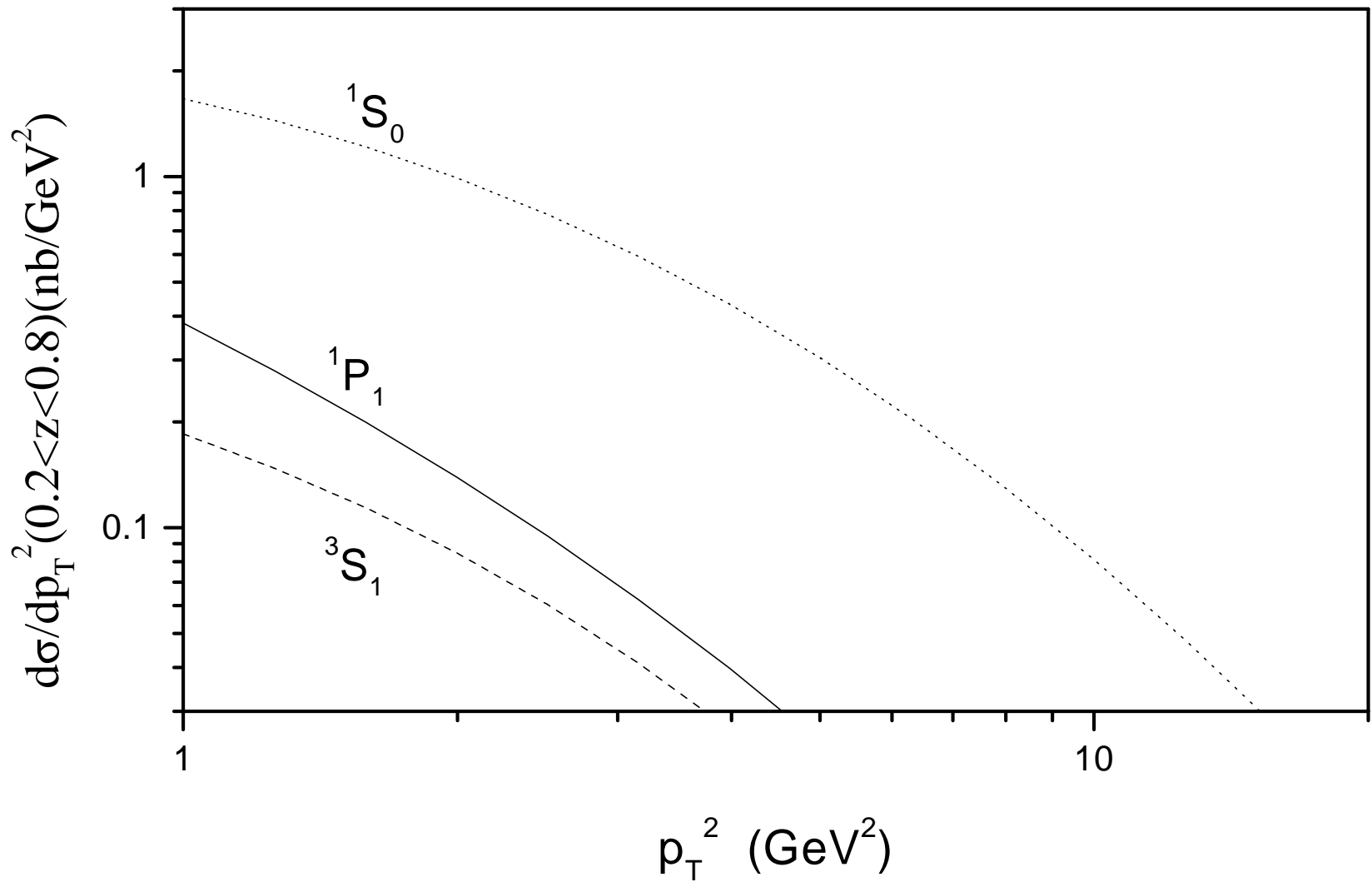
\end{document}